\documentclass[twoside,twocolumn,english]{revtex4}
\usepackage[T1]{fontenc}
\usepackage[latin1]{inputenc}
\usepackage{graphicx}
\usepackage{amssymb}

\makeatletter

\usepackage{babel}
\makeatother
\begin{document}

\title{Signatures of the BCS to Bose crossover in atom shot noise correlations}

\author{Jorge Quintanilla\\
Isis facility, Rutherford Appleton Laboratory, \\
Chilton, Didcot OX11 0QX, Oxfordshire, U.K.}

\begin{abstract}
We discuss the superfluid wave function of an atomic Fermi gas, which
can be measured in an atom shot noise correlation experiment. We note
that it is qualitatively different in the BCS and Bose limits and
on the basis of this we propose quantitative criteria to identify
the crossover between the two regimes.
\end{abstract}
\maketitle
The achievement of quantum degeneracy in Fermi gases \cite{De Marco and Jin},
loading of ultracold atoms on optical lattices \cite{Greiner et al.}
and control of atom-atom interactions using Feshbach resonances suggest
using ultracold atoms as laboratory models of strongly-correlated
electrons. In this respect the recent creation of fermion condensates
\cite{condensates} represents a significant step forward, as it allows
the systematic experimental investigation of one long-standing problem
in that area, namely the crossover from BCS superfluidity of weakly-bound
{}``Cooper pairs'' to Bose-Einstein condensation of tightly-bound
molecules \cite{Randeria}. 

One difficulty is that, although time-of-flight (TOF) experiments
readily yield the momentum distribution of atoms in the condensate,
$n\left(k\right)$, correlation functions are not simple to obtain.
This is in contrast with electrons in solids, for which correlation
functions can be obtained using neutron and X-ray scattering. In this
context it has been proposed to use the pattern of atom shot noise
in TOF images to deduce the density-density correlation function \cite{Altman et al.}.
Recent experiments have implemented these ideas to probe a Bose gas
in the Mott insulating state \cite{Folling et al.} and pairing correlations
in a Fermi gas \cite{Greiner-Regal-Jin}. In the case of a fermion
superfluid, the density-density correlation function can be used to
infer the superfluid wave function \cite{Altman et al.}. Here we
point out that the latter quantity has very different properties in
the BCS and Bose limits and show how to use it to identify the crossover
region. 

We consider a Fermi gas where, as in a recent experiment \cite{Greiner-Regal-Jin},
each atom can have two flavours of the {}``spin''. For simplicity
we assume translational invariance and work at zero temperature. The
quantity of interest is the spin-flip density-density correlation
function between two atoms with different momenta $\hbar\mathbf{k},\hbar\mathbf{k}'$:\[
\Gamma\left(\mathbf{k};\mathbf{k}'\right)=\left\langle \hat{n}_{\uparrow}\left(\mathbf{k}\right)\hat{n}_{\downarrow}\left(\mathbf{k}'\right)\right\rangle -\left\langle \hat{n}_{\uparrow}\left(\mathbf{k}\right)\right\rangle \left\langle \hat{n}_{\downarrow}\left(\mathbf{k}'\right)\right\rangle .\]
Following Altman \emph{et al}.~\cite{Altman et al.} we evaluate
this expression in the BCS ground state, to find\begin{equation}
\Gamma\left(\mathbf{k};\mathbf{k}'\right)=\phi\left(\mathbf{k}\right)\delta_{\mathbf{k},-\mathbf{k}'}.\label{eq:correlations}\end{equation}
 As pointed out in that reference the salient feature is the strong
correlation between particles with opposite momenta, which can be
observed in an experiment of the type described in \cite{Greiner-Regal-Jin},
where the magnitudes $\left|\mathbf{k}\right|$, $\left|\mathbf{k}'\right|$
and the angle $\hat{\mathbf{k}}$ are integrated over and pairing
correlations show up as a peak at $\hat{\mathbf{k}}'=\mathbf{\hat{k}}+\pi$.
Additionally, it was suggested \cite{Altman et al.} to study the
dependence of $\Gamma\left(\mathbf{k},-\mathbf{k}\right)$ on $\left|\mathbf{k}\right|$
by averaging over the other magnitude $\left|\mathbf{k}'\right|$
and the two angles $\hat{\mathbf{k}}$, $\hat{\mathbf{k}}'$. This
yields the $\left|\mathbf{k}\right|$-dependence of the superfluid
wave function, $\phi\left(\mathbf{k}\right).$ Since $\phi\left(\mathbf{k}\right)=\left\langle \hat{c}_{\mathbf{k},\uparrow}^{\dagger}\hat{c}_{-\mathbf{k},\downarrow}^{\dagger}\right\rangle $
it yields essential information on the pairing correlations in the
ground state. It is an experiment of this latter type that we are
considering here.

\begin{figure}
\begin{center}\includegraphics[%
  bb=50bp 40bp 415bp 352bp,
  width=0.45\columnwidth]{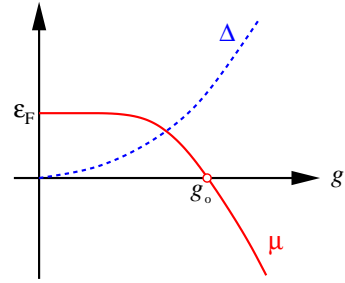}\end{center}

\caption{\label{cap:sketch}Qualitative dependence of the chemical potential,
$\mu$, and amplitude of the pairing potential, $\Delta$, of a fermionic
supefluid as a function of the strength of fermion-fermion attraction,
$g$.}
\end{figure}

Our key assumption is that the BCS ground state, used to derive (\ref{eq:correlations}),
adequately describes, at zero temperature, the correlations present
in both the BCS and Bose limits \cite{Randeria}. Using that ansatz,
it is easy to show that the superfluid wave function is given by \begin{equation}
\phi\left(\mathbf{k}\right)=\frac{\Delta_{\mathbf{k}}}{2\sqrt{\left(\frac{\hbar^{2}k^{2}}{2m}-\mu\right)^{2}+\Delta_{\mathbf{k}}^{2}}},\label{eq:wave-function}\end{equation}
 where $\Delta_{\mathbf{k}}$ is the {}``pairing potential'' of
BCS theory, $m$ the mass of an atom and $\mu$ the chemical potential.
$\mu$ and $\Delta_{\mathbf{k}}$ have to be determined by solving
the BCS self-consistency equations \cite{Tinkham}. Solutions can
be found in the literature for a variety of models with $s$- and
$d$-wave pairing, corresponding to specific choices of a pairing
interaction potential $V\left(\mathbf{k},\mathbf{k}'\right)$--- see,
for example, Refs.~\cite{Andrenacci et al.,Quintanilla et al.,Parish et al.}.
Our emphasis here is on features that do not depend on the model under
consideration.

The qualitative behaviour of the pairing and chemical potentials,
as functions of the strength $g$ of the attractive interaction, is
sketeched in Fig.~\ref{cap:sketch}. For weak attraction, the chemical
potential is independent of $g$ and equal to the Fermi energy, $\varepsilon_{{\rm F}}$,
and the amplitude of the pairing potential, $\Delta$ (where any $\mathbf{k}$-dependence
that does not vary with $g$ has been factored out), grows like $\Delta\sim\exp\left(-1/g\right)$.
As the attraction grows, $\Delta$ continues to grow, albeit more
slowly, and the chemical potential eventually starts to decrease.
This crossover regime sets in smoothly when $\Delta\sim\mu$. In three
dimensions, this happens near the value of $g$ at which the interaction
potential has a bound state and the scattering length diverges, $g_{c}$.
This lowering of the chemical potential corresponds to a shrinking
Fermi surface. %
\footnote{In effect, the spectrum of Bogoliubov quasiparticles is $E_{\mathbf{k}}=\left[\left(\frac{\hbar^{2}k^{2}}{2m}-\mu\right)^{2}+\Delta_{\mathbf{k}}^{2}\right]^{1/2}$
\cite{Tinkham}. Thus the locus, in $\mathbf{k}$-space, of single-particle
excitations with minmum energy is given, along any given direction
$\hat{\mathbf{k}}$, by $\left|\mathbf{k}\right|=\left(2m\mu/\hbar^{2}\right)^{1/2}.$%
} At some value $g=g_{0},$ the chemical potential goes below the bottom
of the band: $\mu<0$. For $g>g_{0}$, the lowest-lying single-particle
excitations occur at $\mathbf{k}=0$ and there is no Fermi surface.
Thus $g_{0}$ is of more physical significance, from the point of
view of the crossover, than $g_{c}$ %
\footnote{In fact if the pairing potential $\Delta_{\mathbf{k}}$ has nodes
in certain directions (as in the case of $d_{x^{2}-y^{2}}$-wave pairing,
relevant to high-temperature superconductors), then the single-particle
excitation spectrum is gapless for $g<g_{0}$ but fully gapped for
$g>g_{0}$ \cite{Randeria-Duan-Shieh}. %
}.

\begin{figure}
\begin{center}\includegraphics[%
  bb=50bp 40bp 300bp 275bp,
  clip,
  width=0.80\columnwidth,
  keepaspectratio]{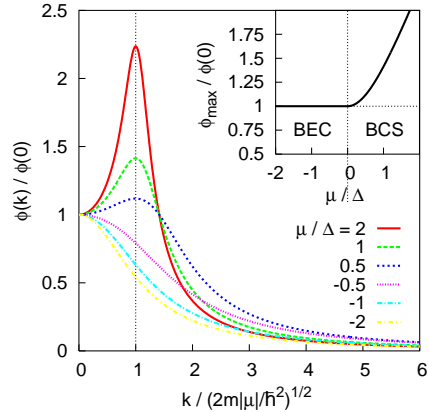}\end{center}

\caption{\label{cap:plot}The superfluid wave function for different values
of $\mu/\Delta,$ as indicated. Inset: dependence of the maximum value
of $\phi\left(k\right)$ on $\mu/\Delta$. }
\end{figure}

We expect the superfluid wave function, Eq.~(\ref{eq:wave-function}),
to reflect the different nature of the BCS and Bose sides of the crossover.
For simplicity we neglect the $\mathbf{k}$-dependence of the pairing
potential: $\Delta_{\mathbf{k}}=\Delta$. Fig.~\ref{cap:plot} shows
$\phi\left(k\right)$ for different values of the ratio $\mu/\Delta$.
The BCS limit corresponds to $\mu\gg\Delta$. In this limit $\phi\left(k\right)$
is maximum on the Fermi surface. As $\mu$ decreases, this surface
becomes smaller so the peak moves to lower values of $k$. For negative
$\mu$, $\phi\left(k\right)$ becomes monotonic, with its maximum
at $k=0$. This makes $\phi\left(k\right)$ qualitatively different
in the BCS regime ($\mu>0$) and the Bose-Einstein regime ($\mu<0$). 

Because the maximum of $\phi\left(k\right)$ is at the locus of lowest-lying
single-particle excitations, tracking the position of this maximum
in $\mathbf{k}$-space allows to monitor the gradual collapse of the
Fermi surface along the BCS to Bose crossover. This can be achieved
by comparing the maximum value of $\phi\left(k\right)$ with its value
at $\mathbf{k}=0$. The ratio is given by\begin{equation}
\frac{\phi_{max}}{\phi\left(0\right)}=\left\{ \begin{array}{cc}
\sqrt{\left(\mu/\Delta\right)^{2}+1} & \textrm{ for }\mu>0\\
1 & \textrm{ for }\mu<0\end{array}\right.\label{eq:maximum}\end{equation}
 This curve is plotted in the inset of Fig.~\ref{cap:plot}. Note
that it depends only on $\mu/\Delta$, not on the specific relation
between $\Delta$ and $\mu$. In that sense Eq.~(\ref{eq:maximum})
is model-independent %
\footnote{The evolution of the supefluid wave function with the strength of
the interaction for a specific model with short-range attraction has
been described in Ref.~\cite{Parish et al.}%
}. 

\begin{figure}
\begin{center}\includegraphics[%
  bb=120bp 100bp 320bp 250bp,
  width=0.70\columnwidth,
  keepaspectratio]{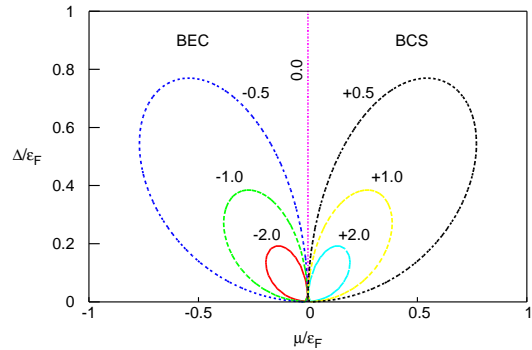}\end{center}

\caption{\label{cap:curvature}Contour plot of the curvature of the superfluid
wave function at $k=0$. The curves correspond to constant values
of $k_{F}^{2}\partial^{2}\phi/\partial k^{2}|_{k=0},$ as indicated.
$k_{{\rm F}}=\left(3\pi^{2}n\right)^{1/3}$ is the Fermi vector, given
in terms of the average density of fermions, $n$, and $\varepsilon_{{\rm F}}=\hbar^{2}k_{F}^{2}/2m$
the Fermi energy.}
\end{figure}

To further quantify the crossover in terms of $\phi\left(k\right)$,
let us calculate its behaviour near $\mathbf{k}=0$. Differentiating
(\ref{eq:wave-function}) we obtain $\partial\phi/\partial k|_{k=0}=0$
and \begin{equation}
\left.\frac{\partial^{2}\phi}{\partial k^{2}}\right|_{k=0}=\frac{\hbar^{2}}{2m}\frac{\mu\Delta}{\left(\Delta^{2}+\mu^{2}\right)^{3/2}}.\label{eq:curvature}\end{equation}
This is plotted in Fig.~\ref{cap:curvature}. We can see that the
curvature of $\phi\left(k\right)$ at $k=0$ changes sign when $\mu$
goes below the bottom of the band. Thus it provides a second test
for the point on the phase diagram where the Fermi surface disappears.
In fact for $\mu>0$ Eqs.~(\ref{eq:maximum}) and (\ref{eq:curvature}),
taken together, completely determine $\mu$ and $\Delta$ from measruable
quantities. We note that the pairing potential can also be obtained
directly from the large-$k$ behaviour of Eq.~(\ref{eq:wave-function}):
$\phi\left(k\right)\approx\left(m\Delta/\hbar^{2}\right)k^{-2}.$

To summarise, we have discussed, on the basis of the BCS ground state,
the superfluid wave function of a fermionic condensate, $\phi\left(k\right)$,
which can be measured directly in shot-noise correlation experiments.
This has quite different behaviour in the BCS and Bose limits, changing
qualitatively when the chemical potential goes below the bottom of
the band. This point in the phase diagram marks the demise of the
Fermi surface. Our main results are Eqs.~(\ref{eq:maximum},\ref{eq:curvature}),
which describe the behaviour of the superfluid wave function along
the BCS to Bose crossover. In particular, they allow to track the
pairing and chemical potentials and to identify where on the phase
diagram $\mu=0$. Our results are independent of any specific model
of the interatomic interaction, relying only on the BCS ansatz for
the ground state.

\textbf{Acknowledgements.} The author would like to thank R. A. Smith
for drawing his attention to noise correlation experiments, C. Hooley,
A. J. Schofield, M. W. Long and N. I. Gidopoulos for useful discussions
and the University of Birmingham for hospitality while some of this
work was carried out.

\textbf{Note added.} Since the author made this work available in
preprint form he has become aware of two unpublished preprints discussing
density correlations in the BCS to Bose crossover: W. Belzig, C. Schroll
and C. Bruder, cond-mat/0412269; and B. Mihaila \emph{et al.}, cond-mat/0502110.

\end{document}